\documentclass[prl,aps,twocolumn,superscriptaddress,showpacs,floatfix]{revtex4}
\usepackage[dvips]{graphicx}
\usepackage{amsmath}
\usepackage{amssymb}
\usepackage[dvips]{graphicx}
\usepackage{amsmath}

\begin{document}

\title{Critical Temperature Curve in the BEC-BCS Crossover}
\author{Evgeni Burovski}
\affiliation{Laboratoire de Physique Th\'{e}orique et Mod\`{e}les Statistiques,
Universit\'{e} Paris-Sud, 91405 Orsay Cedex, France}

\author{Evgeny Kozik}
\affiliation{Department of Physics, University of Massachusetts, Amherst, MA 01003, USA}
\affiliation{Theoretische Physik, ETH Zurich, 8093 Zurich, Switzerland}

\author{Nikolay Prokof'ev}
\affiliation{Department of Physics, University of Massachusetts, Amherst, MA 01003, USA}
\affiliation{Theoretische Physik, ETH Zurich, 8093 Zurich, Switzerland}
\affiliation{Russian Research Center ``Kurchatov Institute'', 123182 Moscow, Russia}

\author{Boris Svistunov}
\affiliation{Department of Physics, University of Massachusetts, Amherst, MA 01003, USA}
\affiliation{Russian Research Center ``Kurchatov Institute'', 123182 Moscow, Russia}

\author{Matthias Troyer}
\affiliation{Theoretische Physik, ETH Zurich, 8093 Zurich, Switzerland}

\begin{abstract}
The strongly-correlated regime of the BCS-BEC crossover can be realized by diluting a
system of two-component fermions with a short-range attractive interaction.
We investigate this system via a novel continuous-space-time diagrammatic determinant
Monte Carlo method and determine the universal curve $T_c/\varepsilon_F$
for the transition temperature between the normal and the superfluid states
as a function of the scattering length with the maximum on the BEC side.
At unitarity, we confirm that $T_c/\varepsilon_F = 0.152(7)$.
\end{abstract}

\maketitle


In the area of ultracold gases, the problem of the crossover
between the Bardeen-Cooper-Schrieffer pairing and the
Bose-Einstein condensation of composite molecules (the so-called
BCS-BEC crossover) has recently received a lot of theoretical and experimental  attention
\cite{Pitaevski-review07}. A dilute
two-component Fermi gas, where the
inter-particle distance is much larger than the interaction range,  features
a remarkable universality at low temperatures. Since the interaction is completely
described by the $s$-wave scattering length $a$, the only
physically relevant coupling parameter is $\kappa=1/k_F a$, where
$k_F$ is the Fermi momentum. One thus obtains a unified and universal description of
systems as diverse as ultracold fermionic gases in magnetic or
optical traps \cite{Pitaevski-review07}, fermions in optical
lattices, inner crusts of neutron stars
\cite{neutron-star_1,GezerlisCarlson2007}, and, plausibly,
excitonic condensates \cite{Franz}.

In the  limit $\kappa \to -\infty$, the Fermi gas is described by the BCS
theory, while for $\kappa \to +\infty$ the fermions pair into
compact bosonic molecules which then form a BEC state below the critical temperature.
Separating these extreme states is a
strongly correlated regime which features the so-called unitary
point $\kappa =0$. At unitarity, the scattering length is infinite and the interaction thus drops out of the
relations between different thermodynamic potentials
making these relations formally identical to those of a
non-interacting Fermi gas \cite{Ho2004}. On the experimental side,
using the technique of a (wide) Feshbach resonance in a system of
cold atoms, one can traverse the whole range of parameter $\kappa$
from the BEC to the BCS limit \cite{Pitaevski-review07}.

Despite considerable recent investigation, the quantitative description of the
BEC-BCS crossover is far from being complete, even for the
simplest case of the equal mixture of two components. Due to the
strongly correlated nature of the problem, analytical mean-field-type calculations (e.g. \cite{classiki,NSR,Zwerger2007}) unavoidably involve approximations, the accuracy of which is difficult to access unless the exact result is known. Renormalization group treatments can be carried out as expansions in either
$\epsilon=4-d$ \cite{Nishida2006}, or $1/N_F$ (where $N_F$ is the
number of fermion species)
\cite{NikolicSachdev2007,Radzihovsky2007}, but the applicability
of these calculations to the physically relevant case of $d=3$ and
$N_F=2$ is not known \textit{a priori}.

Numerical studies of fermionic systems are
computationally demanding and further complicated by the need to study the limit of small
densities to access the universal regime. Some numerical
techniques avoid the fermionic sign problem with a help of
uncontrollable approximations. The restricted path-integral Monte
Carlo (R-PIMC) \cite{Akkineni2007} relies on a variational
ansatz for the nodes of the density-matrix. In the dynamical mean-field
theory (DMFT) approach of Ref.~\cite{Barnea2008}
the physics of extended paired states is reduced to that of a single site
coupled to the self-consistently defined environment. Fortunately,
the unpolarized Fermi gas with contact attraction is an
exceptional case which can be addressed by
sign-problem-free determinant methods without
uncontrollable systematic errors \cite{Bulgac2006, we-short,
we-long}. Moreover, the determinant {\it diagrammatic} MC approach
(DDMC) for lattice fermions \cite{we-long} is completely
free of any systematic error. In simulations of the negative-$U$ Hubbard
model \cite{we-short} with an appropriate extrapolation to zero
filling at the unitary point we previoulsy obtained accurate results for the
critical temperature of the superfluid transition,
$T_c/\varepsilon_F=0.152(7)$, in the units of the Fermi energy
$\varepsilon_F$. This result, however, did not agree with the
estimate obtained by Ref.
\cite{Bulgac2006}  from a visual inspection of the caloric curve,
 using the standard auxiliary field approach \cite{Hirsch}.

\begin{figure}
\includegraphics[width = 0.95\columnwidth,keepaspectratio=true]{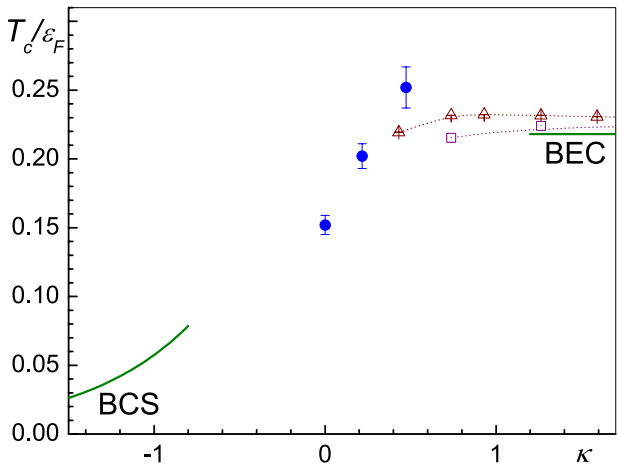}
\caption{(Color online.) The universal results for the critical temperature in the units of the Fermi energy plotted versus $\kappa=1/k_F a$ (circles). The solid lines for negative and positive $\kappa$ represent the limiting behavior of the BCS theory (with the Gorkov-Melik-Barkhudarov correction) and the ideal BEC respectively. For reference, we also plot non-universal results for hard-sphere (triangles) and soft-sphere (squares) bosons Ref.~\cite{Pilati2008}. }
\label{fig:TckFa}
\end{figure}

To efficiently study the critical temperature curve away from the unitarity point and to verify that the final results are model independent -- thereby also resolving the controversy on $T_c/\varepsilon_F$ at unitarity -- we develop a DDMC technique for \textit{continuous} space and time. We can now efficiently simulate models with a simple parabolic dispersion
relation and have completely eliminated lattice
corrections. In this Letter, we first discuss the new scheme and
how to obtain an independent systematic-error-free value for
$T_c/\varepsilon_F$ at unitarity. We are able to reach densities almost $20$ times smaller than those typically accessible with the auxiliary field determinant method \cite{Bulgac2006}, This allows us to perform a reliable extrapolation to the universal limit yielding $T_c/\varepsilon_F =0.152(7)$, in perfect agreement with our previous value \cite{we-short}. Next, we explore the critical temperature at
finite values of $1/k_F a$. Our results, shown in  Fig.~\ref{fig:TckFa}, fix the
general shape of the universal curve $T_c/\varepsilon_F$ versus
$1/k_F a$. The main feature is a substantial maximum of $T_c/\varepsilon_F$ on the BEC side of the crossover.

Our specific model is described by the Hamiltonian
\begin{multline}
 H = \sum_{\sigma=\uparrow,\downarrow} \int\! d\mathbf{x}\, \Psi_\sigma^{\dagger}(\mathbf{x}) \left(
\hat{K} - \mu \right) \Psi_\sigma(\mathbf{x})  \\
 + U \int\! d\mathbf{x}\, \Psi_\uparrow^{\dagger}(\mathbf{x})
\Psi_\downarrow^{\dagger}(\mathbf{x})
\Psi_\downarrow(\mathbf{x}) \Psi_\uparrow(\mathbf{x})\;,
\label{Hamiltonian}
\end{multline}
where $\Psi_\sigma(\mathbf{x})$ is the fermion field operator ($\sigma=\uparrow , \downarrow$), $\mathbf{x}$ is a continuous three-dimensional coordinate, $\mu$
is the chemical potential, $U<0$ is the contact interaction
strength, and $\hat{K}$ is the kinetic energy operator, $\hat{K}
e^{i\mathbf{kx}} = \varepsilon_{\mathbf{k}} e^{i\mathbf{kx}}$,
with $\varepsilon_\mathbf{k}$ being the single-particle dispersion.

The scattering length $a$ is given by the sum of the vacuum
ladder diagrams \cite{LLStatMech2} leading to ($\hbar=1$)
\begin{equation}
\frac{m}{4\pi a}=U^{-1} + \int\! \frac{d \mathbf{k}}{(2 \pi)^3}\frac{1}{2 \varepsilon_\mathbf{k}} \;,
\label{a_sc}
\end{equation}
where $m$ is the fermion mass. For the continuous space model with
$\varepsilon_\mathbf{k} = k^2/2m$ an ultraviolet regularization of
Eq.~\eqref{a_sc} is required. Keeping in mind comparison with Ref.~\cite{Bulgac2006}, where the parabolic dispersion with an ultraviolet cutoff was used, we introduce a microscopic length scale $l_0$ such that
\begin{equation}\label{dispersion}
\varepsilon_\mathbf{k} = \left\{ %
\begin{aligned}
&k^2/2m \;, &k<2\pi/l_0 \; , \\
&\infty \;, &k>2\pi/l_0 \; ,
\end{aligned}
\right.
\end{equation}
yielding
\begin{equation}
m/4\pi a = U^{-1} - U^{-1}_*, \quad\quad U_* = -\pi l_0/m \; . \label{aU}
\end{equation}
%

It is straightforward to generalize the DDMC method for resonant fermions \cite{we-short}  to the continuous model
\eqref{Hamiltonian}. One starts by expanding
the partition function $Z= Tr e^{-\beta H}$, where $\beta=1/k_BT$,
in powers of $U$. The resulting Feynman diagrams consist of
four-point interaction vertices connected by free single-particle
propagators $G^{(0)}_{\sigma}$. A diagram of a given order $p$ is described by the space-time configuration
of the vertices $\mathcal{S}_p = \{(\mathbf{x}_j,\tau_j), \, j=1,\dots,p)\}$ ($\tau \in [0,\beta]$ is the imaginary time) and the topology of propagator lines connecting them
\textit{without} integration over the vertex positions---the latter is done
by the Monte Carlo sampling process. Next, one observes \cite{Rubtsov} that the sum over all topologies
is given by $\det\mathbf{A}^{\uparrow}(\mathcal{S}_p)\det\mathbf{A}^{\downarrow}(\mathcal{S}_p)$,
where $\mathbf{A}^{\sigma}$ is the $p \times p$ matrix,
$A_{ij}^{\sigma}(\mathcal{S}_p) = G^{(0)}_{\sigma}(\mathbf{x}_i - \mathbf{x}_j, \tau_i - \tau_j)$.
In the case of equal densities of the spin components,
the weight of a configuration $\mathcal{S}_p$ is
positive definite:
\begin{equation}
d\mathcal{P}(p, \mathcal{S}_p) = (-U)^p
\left| \det\mathbf{A}(\mathcal{S}_p) \right|^2 \prod_{j=1}^p d\tau_j \, d\mathbf{x}_j \; .
\label{MC_weight}
\end{equation}

The partition function $Z=\sum_{p=0}^{\infty} \int_{\mathcal{S}_p}
d\mathcal{P}$ is calculated stochastically according to the
standard Metropolis-Rosenbluth$^2$-Teller$^2$ algorithm ensuring that
configurations $\mathcal{S}_p$ are generated with the probability density
given by Eg.~(\ref{MC_weight}). The Monte Carlo updates are based on
a worm-algorithm for the four-point correlation function
\cite{we-long}
%
$
G_2(\mathbf{x}, \tau; \mathbf{x}', \tau') = %
\left\langle \mathcal{T}_\tau P(\mathbf{x}, \tau)
P^\dagger(\mathbf{x}', \tau') \right\rangle 
$
%
%
, where $\mathcal{T}_\tau$ indicates time-ordering, $P(\mathbf{x},
\tau) = \Psi_\uparrow(\mathbf{x}, \tau)
\Psi_{\downarrow}(\mathbf{x}, \tau)$ is the pair annihilation
operator, and $\langle \cdots \rangle$ is the thermal average. The
asymptotic value of $\iint \! d\tau d\tau'\, G_2(\mathbf{x}, \tau;
\mathbf{x}', \tau')$ as $|\mathbf{x}-\mathbf{x}'| \to \infty$ is
proportional to the condensate density.

\begin{figure}[htb]
\includegraphics[width = 0.95\columnwidth,keepaspectratio=true]{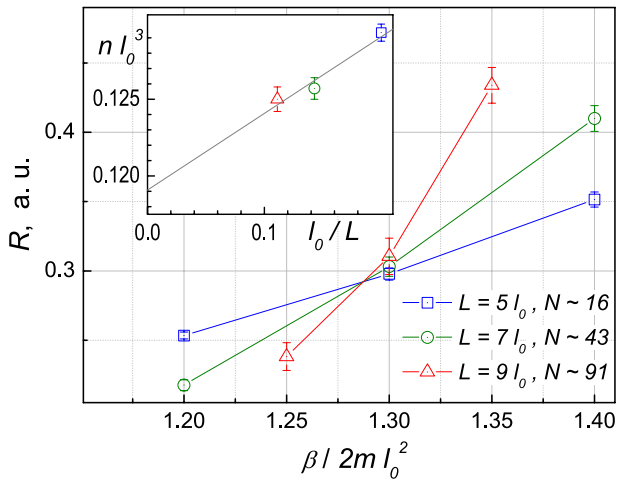}
\caption{(Color online.) Finite-size analysis for $c=0.83$,
$\mu=0.36$, corresponding to $U\approx-7.519$ and $a \approx 1.52$ (in the units of $m=1/2$, $l_0=1$)
yielding $\beta_c=1.290(8)$.
The error bars are one standard deviation and were calculated using the blocking method.
Inset: the thermodynamic limit value of the number density is obtained via a linear fit of $n(L)$ vs. $1/L$. In this case,
$n l_0^3=0.119(2)$, which results in $\zeta=0.492(3)$, $T_c^{(\zeta)}/\varepsilon_F^{(\zeta)}=0.335(6)$, and $\kappa=0.432(3)$. }
\label{fig:crossings}
\end{figure}

Up to statistical errors, the DDMC calculations yield exact results
for a finite system -- in our case a cubic box of a linear size $L$ with
periodic boundary conditions. An efficient way of finding $T_c$
in the thermodynamic limit $L \to \infty$ is to employ the
technique of Binder crossings \cite{Binder} for $R =
L^{1+\eta}\int\,d\mathbf{x}d\mathbf{x}'d\tau d\tau'
G_2(\mathbf{x}, \tau; \mathbf{x}', \tau') /(\beta L^3)^2$ (where
$\eta\approx 0.038$ for the $3$D $U(1)$ universality class), as
discussed in detail in Ref.~\cite{we-long}. It is expected that at
the critical point $R$ becomes scale invariant. By analyzing the
crossings of the family of $R(L,\beta)$ curves one can obtain
$T_c$ with an accuracy of a fraction of percent with a relatively
small number of particles. The thermodynamic limit of the
number density is obtained from a linear extrapolation of
$n(L)$ as a function of $1/L$. An example of
the finite-size analysis for a typical set of parameters is given
in Fig.~\ref{fig:crossings}.

In order to obtain the universal answer for $T_c/ \varepsilon_F$
one finally has to take the limit of $\zeta =n^{1/3}l_0 \to 0$ by
extrapolating numerical data for  $T_c^{(\zeta)}/
\varepsilon_F^{(\zeta)}$ to the dilute limit. We keep $l_0$ constant and
take the limit by lowering the chemical potential $\mu$ and
diluting the system. One can show \cite{we-long} that the
leading-order corrections should be linear in $\zeta \ll 1$:
$T_c^{(\zeta)} / \varepsilon_F^{(\zeta)} = T_c/\varepsilon_F +
\mathrm{const}\times \zeta + o(\zeta)$.
The calculation strategy is as follows: at unitarity ($\kappa\equiv
0$), we fix $U=U_*$ according to Eg.~(\ref{aU}) and perform a
series of simulations for different values of $\mu$, yielding a
set of $T_c^{(\zeta)}/\varepsilon_F^{(\zeta)}$. Then, the
universal value of the critical temperature follows from the
linear extrapolation of $T_c^{(\zeta)}/\varepsilon_F^{(\zeta)}$ to
$\zeta\to 0$.

To obtain $T_c/\varepsilon_F$ away from the resonance, the
procedure has to be modified. Taking the dilute limit for each
value of $\kappa \neq 0$ requires that $a \to \infty$ in such a
way that $1/k_F a$ tends to a fixed finite value $\kappa$. We note that the universal value
of the chemical potential obeys $\mu(T_c)/\varepsilon_F \equiv 2 m
g(\kappa)$, with some function $g(\kappa)$, or, equivalently,
$\lim_{\zeta\to 0} \mu^{(\zeta)}(T_c) a^2 = g(\kappa)/\kappa^2$
implying that for each $\kappa$ one has to keep $\mu a^2 =
\mathrm{const}$. Substituting $a^2=\mu /c$ into \eqref{aU} gives
\begin{equation}
U = U_* \left( 1\pm \frac{mU_*}{4\pi } \sqrt{\frac{\mu}{c}}\, \right)^{-1}
\; ,
\label{u_sigma}
\end{equation}
where the upper (lower) sign corresponds to the BEC side $a>0$ (BCS side
$a<0$). We thus pick a value of $c$ and perform a series of
simulations for smaller and smaller values of $\mu$ with $U$ from Eq.\ \eqref{u_sigma}.
Each simulation yields a finite-$\zeta$ estimate for the
critical temperature $T_c^{(\zeta)}(c)$, density $n^{(\zeta)}(c)$
and $\kappa^{(\zeta)}(c)$. After linear extrapolations to $\zeta\to 0$
we determine the physical value of $T_c/\varepsilon_F$ and the corresponding value of $\kappa$.

\begin{figure}[htb]
\includegraphics[width = 0.95\columnwidth,keepaspectratio=true]{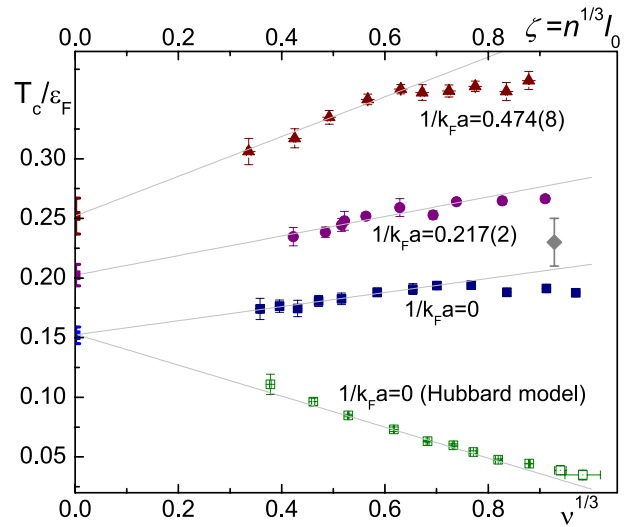}
\caption{(Color online.) The extrapolation of the simulation results to the universal limit $\zeta=n^{1/3} l_0 \to 0$. The procedure yields $T_c/\varepsilon_F=0.152(9)$, $0.202(9)$, and $0.252(15)$ for $\kappa= 1/k_F a = 0$ (squares), $0.217(2)$ (circles), and $0.474(8)$ (triangles) correspondingly. For comparison, we also plot our results for
the Hubbard model (open squares) adapted from Ref.~\cite{we-short}. The estimate of Ref.~\cite{Bulgac2006} at $\kappa=0$
(obtained for finite $\zeta\approx 0.93$) is shown by the diamond. Solid lines are linear fits.}
\label{fig:Tc_dens}
\end{figure}


In Fig.\ \ref{fig:Tc_dens} we show results for the critical temperature as a function of $\zeta$.
For comparison and consistency analysis of $T_c$ at unitarity, we also plot the data for the
Hubbard model \cite{we-short} as a function of the filling factor $\nu$, which plays the same role
as $\zeta$ in the present model.
Note that the non-universal corrections to $T_c/\varepsilon_F$ in $\zeta$ turn out to be positive and
much smaller (at unitarity) than for the Hubbard model. The former fact is important for the simulation efficiency, since the computational complexity of the DDMC technique scales as $(\beta U N)^3$, where
$N$ is the number of fermions, and it is advantageous to simulate at higher temperatures.

It is important to note that at high densities the $T_c^{(\zeta)}/\varepsilon_F^{(\zeta)}$
curves are almost constant and the true asymptotic low-$\zeta$ behavior develops only below $\zeta \approx 0.75$. For a reliable extrapolation it is crucial to vary the density by at least
an order of magnitude, and we did so by diluting the system down to $n \approx 0.04/l_0^3$
($\zeta \approx 0.35$), where we were  limited by the low values of the absolute critical temperature itself. Unfortunately, no dilute-limit extrapolation was performed in Ref.~\cite{Bulgac2006} (their value for $\zeta \approx 0.93$ is shown by the diamond in Fig.~\ref{fig:Tc_dens}).
The total simulation time required to obtain this set of data was approximately $10^6$ CPU hours on Opteron-class workstations.

At unitarity, the extrapolated result for
$T_c^{(\zeta)}/\varepsilon_F^{(\zeta)}$ yields an answer which is
in perfect agreement with $T_c/\varepsilon_F=0.152(7)$ obtained
independently from the $\nu \to 0$ extrapolation of the Hubbard
model data \cite{we-short}. In the latter case, the universal
value is approached from below (see Fig.~\ref{fig:Tc_dens}). This
agreement unambiguously demonstrates that our treatment of
non-universal corrections is reliable in the simulated parameter
range (linear fits for $\zeta < 0.75$). Away from unitarity we
find $T_c/\varepsilon_F=0.202(9)$ and $0.252(15)$ for
$\kappa=0.217(2)$ and $0.474(8)$, respectively.

The results for the strongly correlated regime essentially
determine the general shape of the universal curve
$[T_c/\varepsilon_F](\kappa)$ shown in Fig.~\ref{fig:TckFa}. Deep in
the BEC regime ($\kappa \gg 1$) the critical temperature is that
of a weakly interacting Bose gas of strongly bound dimers which is
expected to increase on approach to the resonance. In the BCS
limit ($\kappa<0$, $|\kappa| \gg 1$) the $T_c$-curve starts from
exponentially small values for $\kappa \to -\infty$, and thus the
crossover between the two limiting regimes necessarily features a
maximum in $T_c/\varepsilon_F$. The results in
Fig.~\ref{fig:TckFa} clearly show that this maximum must be on the
BEC side ($\kappa >0$). The value at the maximum appears to be
surprisingly high. For comparison, we show in Fig.~\ref{fig:TckFa}
the critical temperatures of a Bose gas with  hard- and soft-core
sphere potentials with scattering length $a_B=0.6a$. The
tremendous computational cost required to determine each point in
the crossover regime reliably did not allow us to precisely locate the
position of the maximum in the $k_Fa \sim 1$ region.

The behavior of the critical temperature on the BEC side revealed by our simulations, suggests that the short-range structure of the strongly correlated state is radically different from that of the compact-molecule Bose gas (and, obviously, also from that of the BCS state) in a broad range of $\kappa$. In other words, we are dealing with two crossovers---one is from BCS to the substantial unitarity regime and the other is from the unitarity regime to BEC.

To summarize, we performed first-principle simulations of the
two-component unpolarized Fermi gas with resonant inter-particle
interaction obtaining the universal critical temperature
$T_c/\varepsilon_F=0.152(7)$ at the unitarity point
$\kappa=1/k_Fa=0$ thereby resolving the earlier controversy
between the results of Refs.~\cite{we-long} and \cite{Bulgac2006}.
We also obtain $T_c$ away from unitarity on the BEC side
allowing one to sketch the general dependance
$[T_c/\varepsilon_F](\kappa)$ with a maximum on the BEC side, in a good quantitative agreement with the mean-field-type prediction by Haussmann \textit{et al.} \cite{Zwerger2007}.
After our results were announced
\cite{OurResults}, the Seattle group reconsidered their previous
estimate of the critical temperature \cite{Bulgac2008}. The new
results are in excellent agreement with the values claimed here
both at and away from unitarity.

\acknowledgments{The simulations were performed on the
supercomputers Hreidar at ETH Zurich, Mammoth at the University of
Sherbrooke, Typhon and Athena at the College of Staten Island,
CUNY. The work was supported by the National Science Foundation
under Grant PHY-0653183. E.B. was partially supported by IFRAF. }


\end{document}